# Effect of a new type of healthy and live food supplement on osteoporosis blood parameters and induced rheumatoid arthritis in Wistar rats

Azam Bayat [1, 2, *], Aref Khalkhali [2], Ali Reza Mahjoub[1]

[1]*Department of Chemistry, Tarbiat Modares University, Tehran 14155-4383, Iran*

[1]*NBS organic Company, Istanbul, Turkey*

**Corresponding author: Azam Bayat**

**E-mail address: azam.bayat@modares.ac.ir**

**Abstract**

**Summary** Osteoporosis is a skeletal disorder, characterized by a decrease in bone strength and puts the individual at risk for fracture. On the other hand, rheumatoid arthritis is a systemic disease of unknown etiology that causes inflammation of the joints of the organs.

**Purpose** Due to the destructive effects of these diseases and its increasing prevalence and lack of appropriate medication for treatment, the present study aimed to evaluate the therapeutic effect of a new type of healthy and live food supplement on rheumatoid arthritis and induced osteoporosis in rats.

**Methods** In this research, healthy and live food powder were synthesized by a new and green route. This organic biomaterial was named NBS. The NBS food supplement had various vitamins, macro and micro molecules, and ingredients. The new healthy and nutritious diet showed that the use of this supplement led to the return of the parameters to normal levels.

**Results** The concentration of 12.5 mg/ kg showed the least therapeutic effect and 50 mg/ kg had the highest therapeutic effect for osteoporosis. The results of blood parameters involved in inflammation in both healthy and patient groups showed that the use of complete adjuvant induction causes joint inflammation. In the study of the interaction of the concentrations, it was observed that the concentration of 50 mg/ kg had the highest therapeutic effect against the disease in the studied mice.

**Conclusion** The results showed that the new healthy and viable supplement restores the blood osteoporotic and rheumatoid factors of the mice to normal.

**Introduction**

Osteoporosis is a systemic disease characterized by reduced bone mass and ultrastructural destruction and disruption of bone tissue components and structures [1, 2]. Osteoporosis is the most common bone disease in humans and animals, especially in old age and is more common in females. Osteoporotic fracture statistics show that over time, there will be many problems with the disease [3-5]. From the clinical symptoms, except for a decrease in bone density, there is no other obvious factor for diagnosis. Osteoporosis is divided into primary and secondary forms based on etiology. The primary form may be due to aging or menopause or an unknown cause. However, in the secondary form, different causes can be involved in the disease. These include medications, some hormonal disorders, such as parathyroid hyperplasia, elevated body cortisol levels [6]. Although, it is not definitively treatable after complete establishment, it can be prevented by various methods. One of the predisposing factors for osteoporosis is ovariectomy or complete ovarian resection, in which case the serum estrogen level is decreased and the person becomes susceptible to osteoporosis [7]. In 2003, in a study by Stephan et al. in ovariectomized rats with osteoporosis found that substitution of estrogen in these mice prevented or reduced osteoporosis [8]. A similar study by Takehiko et al. In 2001 found that substituting estrogens prevented bone loss in ovariectomized rats [9]. In 2005, Watkines et al. also presented similar results [10] Getting enough calcium, proper nutrition and physical activity is a good way to prevent osteoporosis during aging.

On the other hand, rheumatoid arthritis is a chronic inflammatory autoimmune disease. Its major pathogenesis occurs in the articular synovium and involves inflammation of the synovial layer of the joint. Over time, the inflamed synovial tissue grows irregularly, forming a tumor-like invasive tumor called panus [11]. The development of autoimmune diseases such as rheumatoid arthritis depends on the interaction of genetic and environmental factors. 50% of the cause of rheumatoid arthritis is genetically dependent [12]. In other cases, environmental factors have been attributed to the disease. Many autoimmune cases do not occur unless an additional incident, such as environmental factors, increases the susceptibility to disease [13]. Although infections can induce autoimmune responses, no specific pathogen has been proven to cause rheumatoid arthritis [14]. The onset of the disease can occur at any age, but the peak age is between 30 and 50 years. Disability in this disease is common and significant. In a group study in the United States, 35 percent of people with rheumatoid arthritis became incapacitated after 10 years [14]. The 2010 American and European Rheumatology Association's diagnostic criteria include joint involvement, acute phase protein inflammation such as CRP, erythrocyte sedimentation rate (ESR), citrulline anti-protein antibody, Rheumatoid factor (RF) and disease duration [15]. Most commonly used medications are mostly used to control joint pain or synovial inflammation and have little effect on the immune-inflammatory currents of the disease, so they cannot prevent the progression of the disease and cartilage and bone deformities in the joints [16].

In this research, healthy and live food supplement were synthesized by a new and green route. This organic biomaterial was named NBS. The NBS healthy and live food powder has various vitamins, macro and micro molecules, and ingredients such as B1, B2, B3, B5, B6, B9, C, K, A, E, D, phosphorus, and etc. The aim of this study was to evaluate serum levels of calcium, phosphorus, magnesium, alkaline phosphatase, albumin and total protein in ovariectomized rats in osteoporosis section. Also, due to the destructive effects of rheumatoid arthritis disease and its increasing prevalence and lack of appropriate medication for treatment of this disease, the present study aimed to evaluate the therapeutic effect of the mentioned healthy and viable food supplement on rheumatoid arthritis.

The healthy and viable drug supplement in the current research may be comparable to chemical supplements. The majority of the multivitamins that are available on the market only meet the needs of the human body. In addition, special attention was paid to their regulation and balance. This highlights the importance of the balanced cellular, molecular, and metabolic function of the human body, which has often been overlooked in other chemical and herbal drugs. In general, emphasis on balance is associated with the improvement and treatment of various diseases.

Another example in this regard is Ganoderma fungi, which has recently been introduced as a therapeutic drug owing to its active compounds for the body, some of which require further investigation. These fungi contain some chemicals that are unknown to the body, including three types of toxins, which may be hazardous to liver health. In addition, the long-term consumption of this material at high doses could lead to adverse complications.

With this background in mind, no comparable foreign and domestic products have been registered that are produced in a similar manner to the processing of cereal grains in the form of a powder supplement for the disease control and treatment.

**Methods**

**Osteoporosis disease**

**Animal Experiments**

For this study, 25 female Wistar rats were selected and divided into 5 groups (n = 5). Feeding and storage conditions were similar for all rats. The rats were kept in special cages and in a bed of pellets at 21-23 ° C for 12 h light and 12 h dark. After a week of getting used to the new situation, a study was started on them. The first group was selected as day zero control and its parameters were considered as normal group. One group was selected as the patient's control group for surgery, in which only ovariectomy was performed to eliminate the effects of surgery as a cause of the error. Three groups were selected as the treatment group on which the ovariectomy was performed. The three groups were maintained for 5 weeks after ovariectomy. At the beginning of the study, group 1 mice, which were the normal group of zero days, were ejected from the sinus area behind the eye after anesthesia with ether. Animals were transferred to the operating room after 3 hours of abstinence. The mice were confined to an ether-containing chamber and were anesthetized using ketamine at 75 mg/ kg and diazepam at 5 mg/ kg intraperitoneally. After settling into a special table, the operation site was shaved and sterilized with 10% betadine. The surgery was performed on the advice of sources from the left and right ventricular approach. After surgical resection, the ovaries were ligated with a two-zero absorbable chromatic ligature and removed along with some surrounding fat and part of the fallopian tubes. Then, the transverse muscles of the abdomen, the internal oblique, and the external oblique at the incision site were sutured using a two-zero absorbable chromic thread in a single individual, and the skin at the incision site was sutured with a single absorbable zero-thread silk thread. After surgery, the rats were treated with gavage of 12.5, 25 and 50 mg/ kg body weight from the aqueous supplement solution for 28 days. In the control group mice only the lateral abdominal area was cut and then sutured. At the end of the period, the rats were anesthetized with ether and blood samples were taken from the sinus region behind the eyes. After blood clotting, the sera were separated by centrifugation at 2500 for 10 min and frozen at -19 ° C. After collecting all samples, all parameters were measured in one day.

**Rheumatoid arthritis disease**

**Animal Experiments**

In this study, 25 males of Wistar male rats with an approximate weight of 200 to 250 g were obtained from Danesh Bonyan Researchers of Green Drug Researchers. Animals were kept at 20 -25 ° C, 12 h light-dark period, and in standard cages. They were transferred to the laboratory one week before the experiment to adapt to the new conditions.

**How to make rheumatoid arthritis**

To induce arthritis on the first day of the experiment, rats were anesthetized with ether, then 0.2 cc FCA was injected into the right knee joint of rats [17]. Symptoms of inflammation were mildly observed from day 1 after injection of the adjuvant in the animal. The animals were randomly divided into 5 groups on day 15.

**First group (normal)**

They did not induce arthritis and received only 10 mg/ kg of solvent (saline) by gavage over days 15-30.

**Group II (Negative Control)**

They induced arthritis and received only physiological serum during the test days.

**Group III (Positive Control)**

Treatment groups 1, 2, and 3 received a dose of 12.5, 25, and 50 mg / kg of healthy dietary supplement after induction of joint inflammation on the first day. On day 30, blood samples were taken directly from the animal's heart. In blood samples, RF, ESR and CRP were measured.

**Results**

Analysis of phenolic compounds in the Nutritional supplement and healthy living (extracted with using HPLC) are shown as Fig. 1. The percent of phenolic compounds of new healthy and live food supplement are Arctigenin 2.34, Gallic Acid 2.41, Quercetin 9.42, Alpha Linoleic Acid 26.80, Linoleic Acid 19.46, Inulin 2.64, Oleic acid 13.24, Palmitic acid 14.98, Stearic acid 3.14 and unknown compound.

**Osteoporosis disease**

Comparison of the results of the both healthy and patient control groups showed that osteoporosis induction was performed correctly and the rats underwent osteoporosis. In the study of blood parameters in the two control groups, a significant difference at the 5% level was observed in the results of the two groups. Comparing the results of the healthy and dietary supplements recipients, it was observed that the use of this supplement led to the return of the parameters to normal levels. In addition, in the study of the interaction of concentrations of healthy and live supplements, it was observed that a concentration of 50 mg / kg had the highest effect and a concentration of 12.5 mg/ kg had the lowest therapeutic effect on osteoporosis. Comparing the results of 50 mg/ kg group of healthy and live dietary supplement with healthy control group, it can be concluded that this supplement eliminates the complications of ovariectomy (Fig. 2 and Table 1).

In this study, in order to evaluate the significance of the data, it is recommended to use ANOVA test. ANOVA test was used to investigate the differences between and within groups. The results of this test showed that there was a significant difference between the groups with 5% probability level. In order to clarify this issue, by Scheffe post hoc test, this significance was tested one by one between groups (Table 2).

**Rheumatoid arthritis disease**

The results of the study of blood parameters involved in inflammation in both healthy and patient control groups showed that the use of complete adjuvant induction causes joint inflammation in the studied mice, so the method of selection to induce joint inflammation is accurate and induced disease in the studied mice. Comparison of the results of the treated groups with the patient control groups showed a significant difference between the studied parameters among these groups. In general, it can be stated that in the study of healthy and live dietary intake groups, the therapeutic effect of this supplement was completely evident, so that the use of this supplement returned the studied factors to normal. The interaction effects of the studied concentrations also showed that the concentration of 50 mg/ kg had the highest therapeutic effect and the concentration of 12.5 mg/ kg had the least therapeutic effect against induced disease in the studied mice (Fig.3 and Table 3).

Meanwhile, in this study, in order to evaluate the significance of the data, it is recommended to use ANOVA test. ANOVA test was used to investigate the differences between and within groups. The results of this test showed that there was a significant difference between the groups with 5% probability level. In order to clarify this issue, by Scheffe post hoc test, this significance was tested one by one between groups (Table 4).

**Discussion**

Osteoporosis is one of the most important diseases that is nowadays considered more popular in human medicine and is very common in men and women especially after menopause. Comparison of the mean of all parameters measured in each group with the control group showed a significant difference ($P < 0.05$). According to the results of this study, calcium levels after ovariectomy showed a significant decrease, which could be due to lack of calcium deposition in bone tissue and renal excretion. These findings are consistent with the results of Calomme et al., which confirm that calcium is reduced after ovariectomy in rats [10]. Kobayashi et al. reported a decrease in serum calcium in rats 4 and 8 weeks after ovariectomy in 2002 [18]. In this study, serum phosphorus levels were significantly increased at 5 weeks after ovariectomy, which may be due to activation of parathyroid hormone and bone resorption of phosphorus with calcium from bone. However, serum phosphorus gradually decreases due to the effect of parathyroid hormone on the kidney and inhibition of phosphorus uptake and urinary secretion and excretion due to calcium uptake by the kidneys [3]. Due to the increase in calcium and phosphorus bone extraction on the one hand, as well as the decrease in their serum levels, due to lack of adsorption and renal excretion of phosphorus, bone mineralization is reduced and the ground for bone fractures is provided. Evaluation of serum alkaline phosphatase mean within and between groups and zero time showed a significant difference between ovariectomized and healthy control groups. It should be noted that in this study the lowest alkaline phosphatase enzyme was in the control group. However, the highest enzyme activity was observed at week 21 after ovariectomy. Due to the activity of osteoclasts in bone and the removal of calcium and phosphorus from bone tissue, the amount of alkaline phosphatase enzyme of bone origin in serum is increased. The results of this study are in line with the findings of other researchers [19].

On the other hand, rheumatoid arthritis is the most common systemic inflammatory disease of the joint and is one of the chronic autoimmune diseases. Its global prevalence has been reported at about one percent. Women, smokers, and people with a positive family history are more likely to develop the disease. Disability in this disease is common and significant. In a cohort study in the United States, 1 in 5 people with rheumatoid arthritis were unable to work after 5 years. Diagnostic criteria include at least one joint with swelling and pain. The likelihood of detecting rheumatoid arthritis increases with increasing number of small joints involved [20]. Chronic synovitis is an inflammatory condition that mostly affects the joints and can damage the cartilage and create bone lesions. Arthritis Rheumatoid arthritis has characteristics that distinguish it from other types of arthritis (inflammation of the joints). Arthritis, for example, is symmetric, meaning that if one knee or hand is involved, the other side is likely to be involved. The joints involved are the metacarpophalangeal, metatarsophalangeal, wrist, and first intermittent joints. In some patients the disease is less severe and is associated with minimal joint damage in a short time, in others progressive and chronic lesions are polyarthritis and cause joint dysfunction [21, 17]. The exact cause of rheumatoid arthritis is unknown, but both cellular and humoral immunity play a role. Among the possible causes of the disease are infections. In this context, bacteria, mycobacteria, mycoplasma and even viruses have been investigated as causative agents. In animal models of rheumatoid arthritis, which is dependent on TLR2, TLR4, and TLR9, rodents were injected with the SCW streptococcal cell wall and it was observed that severe acute polyarthritis developed, which was then improved by a chronic T cell-dependent polyarthritis. It is very similar to rheumatoid arthritis [22]. Purulent sore throat as a causative agent, as shown in rheumatic fever, has not been proven in rheumatoid arthritis, however, in a special subgroup of rheumatoid arthritis called steel disease, this relationship begins with adulthood. In the diagnostic criteria for the disease in 1987 by Cush Farangit was not one of the criteria, and in the 1992 criteria set by Yamaguchi, Farangit was the minor criterion, and in the 2008 Fautrel criteria, Farangit was mentioned as the major criterion in the diagnosis of this disease [23].

**Conclusion**

The results of this study show that the new healthy and viable dietary supplement restores the blood osteoporotic factors to normal and improves the disease in mice. The most appropriate therapeutic dose of this food is a concentration of 50 mg / kg. Based on the research findings, it can be stated that a healthy and viable dietary supplement can be used as a drug to reduce the symptoms of osteoporosis, or more properly, as a drug to treat osteoporosis. The results of this study also showed that treatment with this healthy and live food supplement would restore the blood rheumatoid factors of the studied mice to normal and eliminate the symptoms of rheumatoid arthritis. Healthy and nutritious dietary supplements can be used to treat joint inflammation.

**Acknowledgements** The authors acknowledge finical support of NBS Organic Company and Tarbiat Modares University for supporting this work.

**Funding information** None.

**Conflicts of interest**

Azam Bayat, Aref Khalkhali, and Ali Reza Mahjoub declare that they have no conflict of interest.

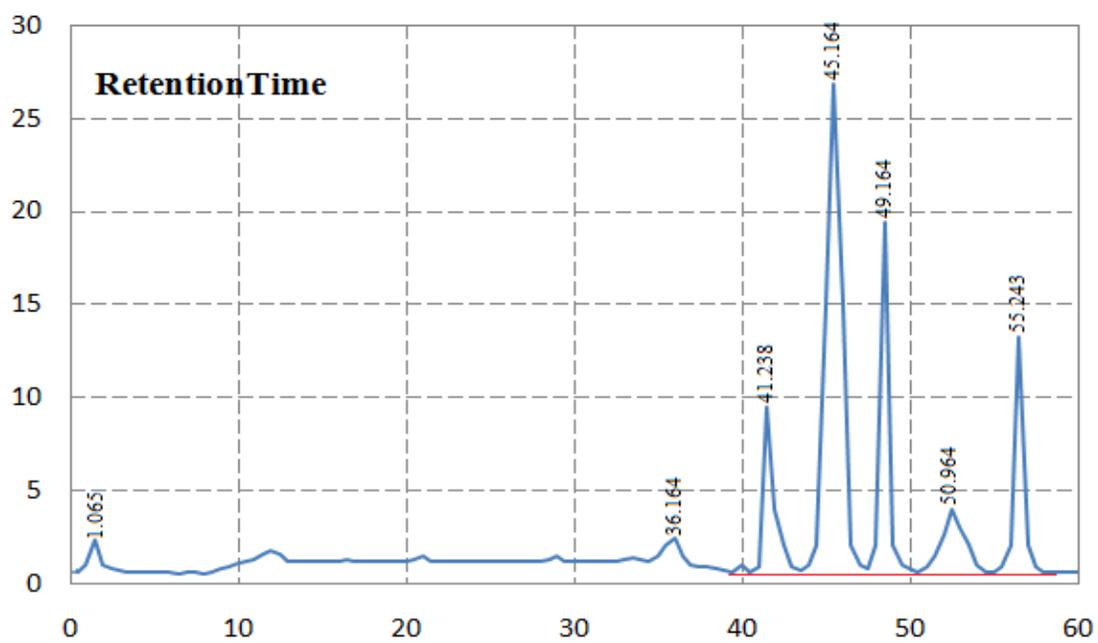

**Fig. 1.** Analysis of phenolic compounds in the Nutritional supplement and healthy living (extracted with using HPLC)

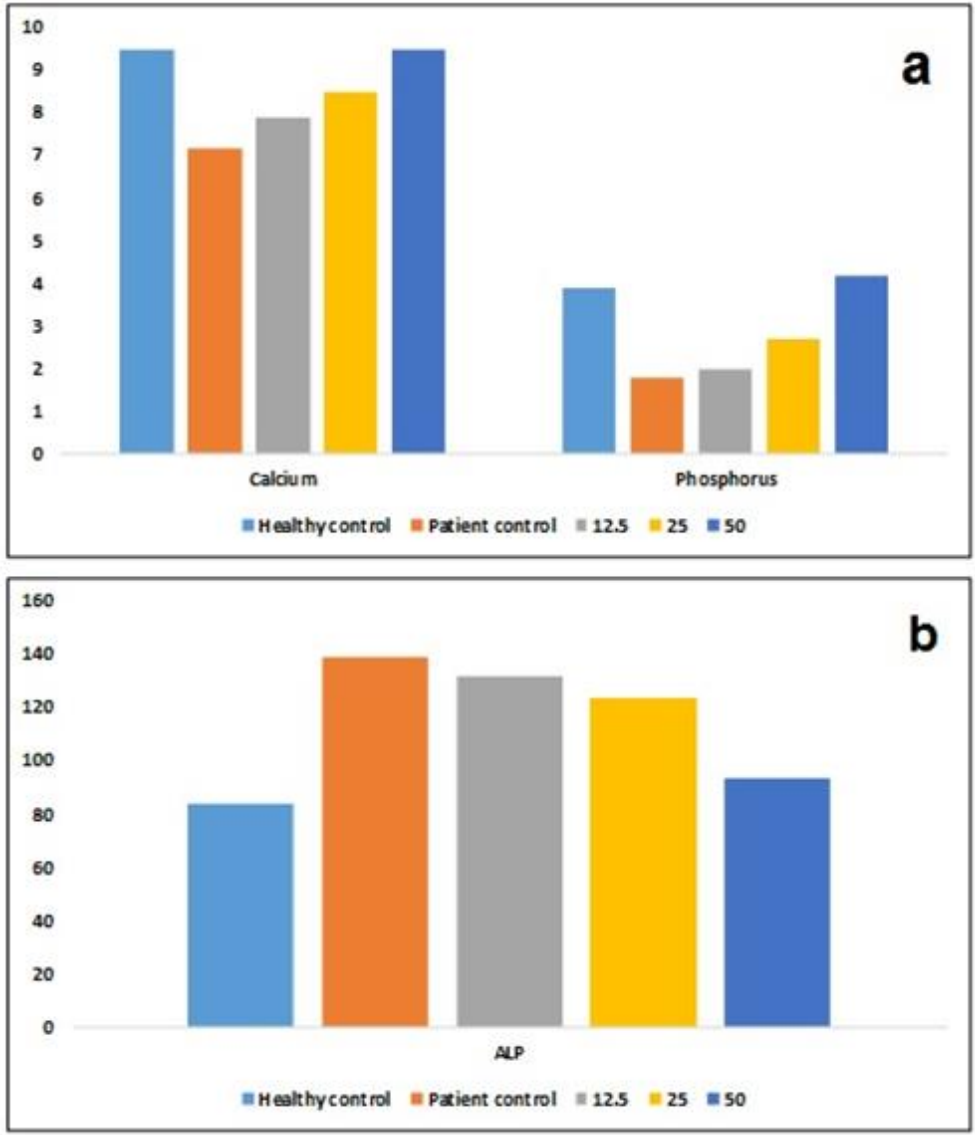

**Fig. 2.** Comparison of the results of the serum parameters (a) Calcium and Phosphorus, and (b) ALP of mice in control groups and treatment with healthy and live food supplements.

| Analysis a | Result | | | | | Range |
|---|---|---|---|---|---|---|
| | Mice (1) | Mice (2) | Mice (3) | Mice (4) | Mice (5) | |
| Calcium | 9.7 | 9.1 | 9.6 | 10.2 | 9.4 | 8.9-10.01 mg/dl |
| Phosphorus | 3.7 | 3.9 | 4.6 | 4.4 | 3.6 | 2.5-4.5 mg/dl |
| ALP | 83 | 79 | 92 | 84 | 98 | 40-129 IU/L |

| Analysis b | Result | | | | | Range |
|---|---|---|---|---|---|---|
| | Mice (1) | Mice (2) | Mice (3) | Mice (4) | Mice (5) | |
| Calcium | 7.3 | 6.5 | 7.4 | 8.1 | 7.2 | 8.9-10.01 mg/dl |
| Phosphorus | 1.2 | 1.7 | 2.4 | 1.3 | 1.8 | 2.5-4.5 mg/dl |
| ALP | 164 | 127 | 139 | 152 | 143 | 40-129 IU/L |

| Analysis c | Result | | | | | Range |
|---|---|---|---|---|---|---|
| | Mice (1) | Mice (2) | Mice (3) | Mice (4) | Mice (5) | |
| Calcium | 8.2 | 8.4 | 7.7 | 7.9 | 7.6 | 8.9-10.01 mg/dl |
| Phosphorus | 1.9 | 2.3 | 2.6 | 1.8 | 2.1 | 2.5-4.5 mg/dl |
| ALP | 133 | 141 | 129 | 138 | 134 | 40-129 IU/L |

| Analysis | Result | | | | | Range |
|---|---|---|---|---|---|---|
| d | Mice (1) | Mice (2) | Mice (3) | Mice (4) | Mice (5) | |
| Calcium | 8.3 | 8.7 | 8.3 | 8.6 | 8.4 | 8.9-10.01 mg/dl |
| Phosphorus | 2.7 | 2.8 | 2.5 | 2.8 | 2.6 | 2.5-4.5 mg/dl |
| ALP | 119 | 121 | 128 | 124 | 126 | 40-129 IU/L |

| Analysis | Result | | | | | Range |
|---|---|---|---|---|---|---|
| e | Mice (1) | Mice (2) | Mice (3) | Mice (4) | Mice (5) | |
| Calcium | 9.5 | 10.1 | 9.2 | 8.7 | 9.6 | 8.9-10.01 mg/dl |
| Phosphorus | 4.4 | 3.6 | 4.2 | 3.4 | 3.7 | 2.5-4.5 mg/dl |
| ALP | 86 | 98 | 112 | 94 | 102 | 40-129 IU/L |

**Table 1.** Results of the osteoporosis blood parameters measurement in (a) healthy control rats, (b) control group rats without medication, (c) healthy live-supplement group at a dose of 12.5, (d) healthy live-supplement group at a dose of 25, and (e) healthy live-supplement group at a dose of 50.

Calcium

| (I) VAR00001 | (J) VAR00001 | Mean Difference (I-J) | Std. Error | Sig. | 95% Confidence Interval | |
|---|---|---|---|---|---|---|
| | | | | | Lower Bound | Upper Bound |
| Healthy group | Patient group | 2.3000* | .2688 | .000 | 1.390 | 3.210 |
| | Concentration 12.5 | 1.6400* | .2688 | .000 | .730 | 2.550 |
| | Concentration 25 | 1.1400* | .2688 | .009 | .230 | 2.050 |
| | Concentration 50 | .1800 | .2688 | .977 | -.730 | 1.090 |
| Patient group | Healthy group | -2.3000* | .2688 | .000 | -3.210 | -1.390 |
| | Concentration 12.5 | -.6600 | .2688 | .238 | -1.570 | .250 |
| | Concentration 25 | -1.1600* | .2688 | .008 | -2.070 | -.250 |
| | Concentration 50 | -2.1200* | .2688 | .000 | -3.030 | -1.210 |
| Concentration 12.5 | Healthy group | -1.6400* | .2688 | .000 | -2.550 | -.730 |
| | Patient group | .6600 | .2688 | .238 | -.250 | 1.570 |
| | Concentration 25 | -.5000 | .2688 | .502 | -1.410 | .410 |
| | Concentration 50 | -1.4600* | .2688 | .001 | -2.370 | -.550 |
| Concentration 25 | Healthy group | -1.1400* | .2688 | .009 | -2.050 | -.230 |
| | Patient group | 1.1600* | .2688 | .008 | .250 | 2.070 |
| | Concentration 12.5 | .5000 | .2688 | .502 | -.410 | 1.410 |
| | Concentration 50 | -.9600* | .2688 | .035 | -1.870 | -.050 |
| Concentration 50 | Healthy group | -.1800 | .2688 | .977 | -1.090 | .730 |
| | Patient group | 2.1200* | .2688 | .000 | 1.210 | 3.030 |
| | Concentration 12.5 | 1.4600* | .2688 | .001 | .550 | 2.370 |
| | Concentration 25 | .9600* | .2688 | .035 | .050 | 1.870 |

*. The mean difference is significant at the 0.05 level.

Phosphorus

| (I) VAR00001 | (J) VAR00001 | Mean Difference (I-J) | Std. Error | Sig. | 95% Confidence Interval | |
|---|---|---|---|---|---|---|
| | | | | | Lower Bound | Upper Bound |
| Healthy group | Patient group | 2.3600* | .2397 | .000 | 1.549 | 3.171 |
| | Concentration 12.5 | 1.9000* | .2397 | .000 | 1.089 | 2.711 |
| | Concentration 25 | 1.3600* | .2397 | .000 | .549 | 2.171 |
| | Concentration 50 | .1800 | .2397 | .965 | -.631 | .991 |
| Patient group | Healthy group | -2.3600* | .2397 | .000 | -3.171 | -1.549 |
| | Concentration 12.5 | -.4600 | .2397 | .471 | -1.271 | .351 |
| | Concentration 25 | -1.0000* | .2397 | .011 | -1.811 | -.189 |
| | Concentration 50 | -2.1800* | .2397 | .000 | -2.991 | -1.369 |
| Concentration 12.5 | Healthy group | -1.9000* | .2397 | .000 | -2.711 | -1.089 |
| | Patient group | .4600 | .2397 | .471 | -.351 | 1.271 |
| | Concentration 25 | -.5400 | .2397 | .315 | -1.351 | .271 |
| | Concentration 50 | -1.7200* | .2397 | .000 | -2.531 | -.909 |
| Concentration 25 | Healthy group | -1.3600* | .2397 | .000 | -2.171 | -.549 |
| | Patient group | 1.0000* | .2397 | .011 | .189 | 1.811 |
| | Concentration 12.5 | .5400 | .2397 | .315 | -.271 | 1.351 |
| | Concentration 50 | -1.1800* | .2397 | .002 | -1.991 | -.369 |
| Concentration 50 | Healthy group | -.1800 | .2397 | .965 | -.991 | .631 |
| | Patient group | 2.1800* | .2397 | .000 | 1.369 | 2.991 |
| | Concentration 12.5 | 1.7200* | .2397 | .000 | .909 | 2.531 |
| | Concentration 25 | 1.1800* | .2397 | .002 | .369 | 1.991 |

*. The mean difference is significant at the 0.05 level.

ALP

| (I) VAR00001 | (J) VAR00001 | Mean Difference (I-J) | Std. Error | Sig. | 95% Confidence Interval | |
|---|---|---|---|---|---|---|
| | | | | | Lower Bound | Upper Bound |
| Healthy group | Patient group | -57.800* | 5.512 | .000 | -76.46 | -39.14 |
| | Concentration 12.5 | -47.800* | 5.512 | .000 | -66.46 | -29.14 |
| | Concentration 25 | -36.400* | 5.512 | .000 | -55.06 | -17.74 |
| | Concentration 50 | -11.200 | 5.512 | .415 | -29.86 | 7.46 |
| Patient group | Healthy group | 57.800* | 5.512 | .000 | 39.14 | 76.46 |
| | Concentration 12.5 | 10.000 | 5.512 | .526 | -8.66 | 28.66 |
| | Concentration 25 | 21.400* | 5.512 | .019 | 2.74 | 40.06 |
| | Concentration 50 | 46.600* | 5.512 | .000 | 27.94 | 65.26 |
| Concentration 12.5 | Healthy group | 47.800* | 5.512 | .000 | 29.14 | 66.46 |
| | Patient group | -10.000 | 5.512 | .526 | -28.66 | 8.66 |
| | Concentration 25 | 11.400 | 5.512 | .398 | -7.26 | 30.06 |
| | Concentration 50 | 36.600* | 5.512 | .000 | 17.94 | 55.26 |
| Concentration 25 | Healthy group | 36.400* | 5.512 | .000 | 17.74 | 55.06 |
| | Patient group | -21.400* | 5.512 | .019 | -40.06 | -2.74 |
| | Concentration 12.5 | -11.400 | 5.512 | .398 | -30.06 | 7.26 |
| | Concentration 50 | 25.200* | 5.512 | .005 | 6.54 | 43.86 |
| Concentration 50 | Healthy group | 11.200 | 5.512 | .415 | -7.46 | 29.86 |
| | Patient group | -46.600* | 5.512 | .000 | -65.26 | -27.94 |
| | Concentration 12.5 | -36.600* | 5.512 | .000 | -55.26 | -17.94 |
| | Concentration 25 | -25.200* | 5.512 | .005 | -43.86 | -6.54 |

*. The mean difference is significant at the 0.05 level.

**Table 2.** Results of ANOVA analysis and Scheffe post hoc test data for calcium, phosphorus and alkaline phosphatase results of the studied mice.

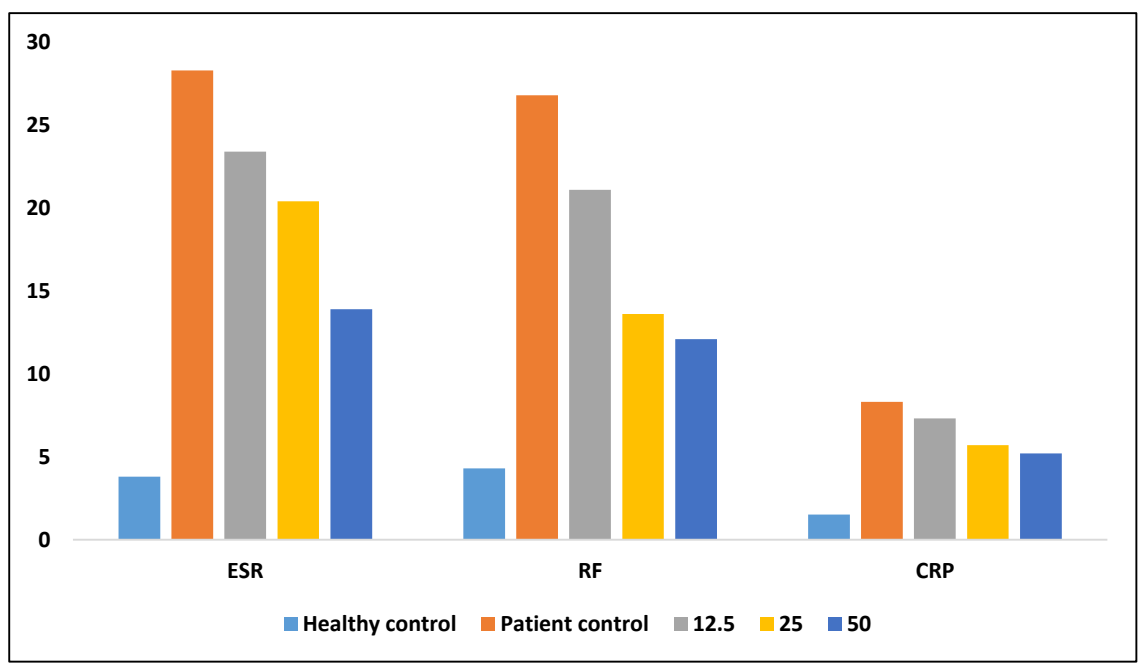

**Fig. 3.** Comparison of rheumatoid arthritis parameters of mice in control groups and treatment with healthy and live supplements.

| Analysis **a** | Result | | | | | Range |
|---|---|---|---|---|---|---|
| | Mice (1) | Mice (2) | Mice (3) | Mice (4) | Mice (5) | |
| ESR | 3.6 | 4.9 | 12.3 | 8.4 | 5.1 | Negative ≤ 22 |
| RF | 6.4 | 2.7 | 1.8 | 5.9 | 3.1 | Negative ≤ 20 |
| CRP | 1.7 | 1.6 | 1.2 | 2.4 | 1.8 | Negative ≤ 6 |

| Analysis **b** | Result | | | | | Range |
|---|---|---|---|---|---|---|
| | Mice (1) | Mice (2) | Mice (3) | Mice (4) | Mice (5) | |
| ESR | 29.4 | 31.2 | 27.2 | 29.4 | 26.8 | Negative ≤ 22 |
| RF | 27.2 | 23.9 | 26.8 | 24.5 | 23.4 | Negative ≤ 20 |
| CRP | 7.9 | 8.6 | 8.2 | 7.6 | 8.4 | Negative ≤ 6 |

| Analysis **c** | Result | | | | | Range |
|---|---|---|---|---|---|---|
| | Mice (1) | Mice (2) | Mice (3) | Mice (4) | Mice (5) | |
| ESR | 22.3 | 21.9 | 26.1 | 23.4 | 22.8 | Negative ≤ 22 |
| RF | 16.2 | 18.7 | 20.1 | 18.4 | 19.6 | Negative ≤ 20 |
| CRP | 6.9 | 6.6 | 8.1 | 7.4 | 6.2 | Negative ≤ 6 |

| Analysis | Result | | | | | Range |
|---|---|---|---|---|---|---|
| d | Mice (1) | Mice (2) | Mice (3) | Mice (4) | Mice (5) | |
| ESR | 20.7 | 19.9 | 20.1 | 20.6 | 20.4 | Negative ≤ 22 |
| RF | 16.9 | 16.3 | 17.4 | 16.8 | 16.2 | Negative ≤ 20 |
| CRP | 4.9 | 5.4 | 5.3 | 6.1 | 5.9 | Negative ≤ 6 |

| Analysis | Result | | | | | Range |
|---|---|---|---|---|---|---|
| e | Mice (1) | Mice (2) | Mice (3) | Mice (4) | Mice (5) | |
| ESR | 14.6 | 12.9 | 15.2 | 14.3 | 16.4 | Negative ≤ 22 |
| RF | 12.3 | 11.7 | 11.8 | 13.6 | 11.2 | Negative ≤ 20 |
| CRP | 5.2 | 5.8 | 5.4 | 5.6 | 5.1 | Negative ≤ 6 |

**Table 3**. Results of blood factors related to joint inflammation in (a) healthy control group, (b) patients control group, (c) rats receiving a healthy diet with a concentration of 12.5, (d) rats receiving a healthy diet with a concentration of 25, and (e) rats receiving a healthy diet with a concentration of 50.

ESR

| (I) VAR00001 | (J) VAR00001 | Mean Difference (I-J) | Std. Error | Sig. | 95% Confidence Interval | |
|---|---|---|---|---|---|---|
| | | | | | Lower Bound | Upper Bound |
| Healthy control | Patient control | -21.94000* | 1.26990 | .000 | -26.2397 | -17.6403 |
| | NBS 12.5 | -16.44000* | 1.26990 | .000 | -20.7397 | -12.1403 |
| | NBS 25 | -13.48000* | 1.26990 | .000 | -17.7797 | -9.1803 |
| | NBS 50 | -7.82000* | 1.26990 | .000 | -12.1197 | -3.5203 |
| Patient control | Healthy control | 21.94000* | 1.26990 | .000 | 17.6403 | 26.2397 |
| | NBS 12.5 | 5.50000* | 1.26990 | .008 | 1.2003 | 9.7997 |
| | NBS 25 | 8.46000* | 1.26990 | .000 | 4.1603 | 12.7597 |
| | NBS 50 | 14.12000* | 1.26990 | .000 | 9.8203 | 18.4197 |
| NBS 12.5 | Healthy control | 16.44000* | 1.26990 | .000 | 12.1403 | 20.7397 |
| | Patient control | -5.50000* | 1.26990 | .008 | -9.7997 | -1.2003 |
| | NBS 25 | 2.96000 | 1.26990 | .284 | -1.3397 | 7.2597 |
| | NBS 50 | 8.62000* | 1.26990 | .000 | 4.3203 | 12.9197 |
| NBS 25 | Healthy control | 13.48000* | 1.26990 | .000 | 9.1803 | 17.7797 |
| | Patient control | -8.46000* | 1.26990 | .000 | -12.7597 | -4.1603 |
| | NBS 12.5 | -2.96000 | 1.26990 | .284 | -7.2597 | 1.3397 |
| | NBS 50 | 5.66000* | 1.26990 | .006 | 1.3603 | 9.9597 |
| NBS 50 | Healthy control | 7.82000* | 1.26990 | .000 | 3.5203 | 12.1197 |
| | Patient control | -14.12000* | 1.26990 | .000 | -18.4197 | -9.8203 |
| | NBS 12.5 | -8.62000* | 1.26990 | .000 | -12.9197 | -4.3203 |
| | NBS 25 | -5.66000* | 1.26990 | .006 | -9.9597 | -1.3603 |

*. The mean difference is significant at the 0.05 level.

RF

| (I) VAR00001 | (J) VAR00001 | Mean Difference (I-J) | Std. Error | Sig. | 95% Confidence Interval | |
|---|---|---|---|---|---|---|
| | | | | | Lower Bound | Upper Bound |
| Healthy control | Patient control | -21.18000* | .91691 | .000 | -24.2846 | -18.0754 |
| | NBS 12.5 | -14.62000* | .91691 | .000 | -17.7246 | -11.5154 |
| | NBS 25 | -12.74000* | .91691 | .000 | -15.8446 | -9.6354 |
| | NBS 50 | -8.14000* | .91691 | .000 | -11.2446 | -5.0354 |
| Patient control | Healthy control | 21.18000* | .91691 | .000 | 18.0754 | 24.2846 |
| | NBS 12.5 | 6.56000* | .91691 | .000 | 3.4554 | 9.6646 |
| | NBS 25 | 8.44000* | .91691 | .000 | 5.3354 | 11.5446 |
| | NBS 50 | 13.04000* | .91691 | .000 | 9.9354 | 16.1446 |
| NBS 12.5 | Healthy control | 14.62000* | .91691 | .000 | 11.5154 | 17.7246 |
| | Patient control | -6.56000* | .91691 | .000 | -9.6646 | -3.4554 |
| | NBS 25 | 1.88000 | .91691 | .406 | -1.2246 | 4.9846 |
| | NBS 50 | 6.48000* | .91691 | .000 | 3.3754 | 9.5846 |
| NBS 25 | Healthy control | 12.74000* | .91691 | .000 | 9.6354 | 15.8446 |
| | Patient control | -8.44000* | .91691 | .000 | -11.5446 | -5.3354 |
| | NBS 12.5 | -1.88000 | .91691 | .406 | -4.9846 | 1.2246 |
| | NBS 50 | 4.60000* | .91691 | .002 | 1.4954 | 7.7046 |
| NBS 50 | Healthy control | 8.14000* | .91691 | .000 | 5.0354 | 11.2446 |
| | Patient control | -13.04000* | .91691 | .000 | -16.1446 | -9.9354 |
| | NBS 12.5 | -6.48000* | .91691 | .000 | -9.5846 | -3.3754 |
| | NBS 25 | -4.60000* | .91691 | .002 | -7.7046 | -1.4954 |

*. The mean difference is significant at the 0.05 level.

CRP

| (I) VAR00001 | (J) VAR00001 | Mean Difference (I-J) | Std. Error | Sig. | 95% Confidence Interval | |
|---|---|---|---|---|---|---|
| | | | | | Lower Bound | Upper Bound |
| Healthy control | Patient control | -6.40000* | .31023 | .000 | -7.4504 | -5.3496 |
| | NBS 12.5 | -5.30000* | .31023 | .000 | -6.3504 | -4.2496 |
| | NBS 25 | -3.78000* | .31023 | .000 | -4.8304 | -2.7296 |
| | NBS 50 | -3.68000* | .31023 | .000 | -4.7304 | -2.6296 |
| Patient control | Healthy control | 6.40000* | .31023 | .000 | 5.3496 | 7.4504 |
| | NBS 12.5 | 1.10000* | .31023 | .037 | .0496 | 2.1504 |
| | NBS 25 | 2.62000* | .31023 | .000 | 1.5696 | 3.6704 |
| | NBS 50 | 2.72000* | .31023 | .000 | 1.6696 | 3.7704 |
| NBS 12.5 | Healthy control | 5.30000* | .31023 | .000 | 4.2496 | 6.3504 |
| | Patient control | -1.10000* | .31023 | .037 | -2.1504 | -.0496 |
| | NBS 25 | 1.52000* | .31023 | .002 | .4696 | 2.5704 |
| | NBS 50 | 1.62000* | .31023 | .001 | .5696 | 2.6704 |
| NBS 25 | Healthy control | 3.78000* | .31023 | .000 | 2.7296 | 4.8304 |
| | Patient control | -2.62000* | .31023 | .000 | -3.6704 | -1.5696 |
| | NBS 12.5 | -1.52000* | .31023 | .002 | -2.5704 | -.4696 |
| | NBS 50 | .10000 | .31023 | .999 | -.9504 | 1.1504 |
| NBS 50 | Healthy control | 3.68000* | .31023 | .000 | 2.6296 | 4.7304 |
| | Patient control | -2.72000* | .31023 | .000 | -3.7704 | -1.6696 |
| | NBS 12.5 | -1.62000* | .31023 | .001 | -2.6704 | -.5696 |
| | NBS 25 | -.10000 | .31023 | .999 | -1.1504 | .9504 |

*. The mean difference is significant at the 0.05 level.

**Table 4.** Results of ANOVA analysis and Scheffe post hoc test results for ESR, RF and CRP results of the studied mice.